\documentclass[showpacs,preprintnumbers,amsmath,amssymb,12pt]{article}
\usepackage{graphicx,amsfonts,amssymb,amsthm,amsmath,graphics,epsfig,pstricks,fancyhdr,fancybox}
\usepackage{dcolumn}
\usepackage{bm}

\newcommand{\bea}{\begin{eqnarray}}
\newcommand{\eea}{\end{eqnarray}}
\newcommand{\be}{\begin{equation}}
\newcommand{\ee}{\end{equation}}
\newcommand{\refeq}[1]{~(\ref{#1})}

\newcommand{\EXP}[1]{\bm{E}\left[{#1}\right]}

\begin{document}
\thispagestyle{empty}

\title{\LARGE \textbf{Logistic and $\theta$-logistic models\\ in population dynamics: \\
General analysis and exact results}}

\author{
Nicola \textsc{Cufaro Petroni}\footnote{cufaro@ba.infn.it}  \\
Dipartimento di \textsl{Matematica} and \textsl{TIRES}, University
of Bari (\textit{Ret})\\
 \vspace{7pt}
and \textsl{INFN} Sezione di Bari; via E. Orabona 4, 70125 Bari,
Italy\\
Salvatore \textsc{De Martino}\footnote{demartino@sa.infn.it}\\
Dipartimento di \textsl{Ingegneria dell'Informazione ed Elettrica}\\
\textsl{e Matematica applicata}, University of Salerno (\textit{Ret})\\
and INFN, Sezione di Napoli, Gruppo Collegato di Salerno \\
 \vspace{7pt}
via Giovanni Paolo II, 132, 84084 Fisciano (SA), Italy\\
Silvio \textsc{De Siena}\footnote{silvio.desiena@gmail.com}\\
Dipartimento di \textsl{Ingegneria Industriale}, University of Salerno (\textit{Ret})\\
personal address: via Bastioni 15, 84122 Salerno, Italy }

\date{}

\maketitle

\vspace{-1cm}

\begin{abstract}
\noindent In the present paper we provide the closed form of the
path-like solutions for the logistic and $\theta$-logistic
stochastic differential equations, along with the exact expressions
of both their probability density functions and their moments. We
simulate in addition a few typical sample trajectories, and we
provide a few examples of numerical computation of the said closed
formulas at different noise intensities: this shows in particular
that an increasing randomness -- while making the process more
unpredictable -- asymptotically tends to suppress in average the
logistic growth. These main results are preceded by a discussion of
the noiseless, deterministic versions of these models: a prologue
which turns out to be instrumental -- on the basis of a few
simplified but functional hypotheses -- to frame the logistic and
$\theta$-logistic equations in a unified context, within which also
the Gompertz model emerges from an anomalous scaling.
\end{abstract}

\noindent \textsc{Keywords}: Population dynamics; Logistic
equations; Stochastic growth models

\section{Introduction}
\label{Intro}

Investigation of population dynamics can be traced back to the
Fibonacci series in thirteenth century, and have been then developed
until the present day \cite{muller,Ovaskainen,Salisbury} with the
introduction of various models designed to describe a very large
number of systems with both theoretical and practical relevance
\cite{Murray1,Murray2}. Phenomenological equations have been
proposed to account for the macroscopic behaviors resulting from a
suitable averaging.

On a macroscopic level, two approaches became very popular along the
years and can now be considered as prototypical: the Verhulst
(logistic) model \cite{Verhulst} and the Gompertz model
\cite{Gompertz}, both introduced in the first half of the nineteenth
century, and then resumed and developed in the first half of the
twentieth century. The $\theta$-logistic equation (Richards Model)
\cite{Richards,GilpinAyala} was subsequently added as a flexible
generalization of the logistic evolution. The corresponding laws can
indeed be obtained resorting to a proportionality between the
differential increment of the size of a system and its current size,
and then suitably correcting it by adding a nonlinear factor that
prevents an un-physical (Malthusian) explosion allowed only in the
first stage of the evolution: this will eventually drive the system
toward a finite asymptotic dimension, namely to a stable equilibrium
point. The said correction is in fact related to the finite amount
of resources available for a given system, and to its growing
density, two features both leading to a reduction of the resources
allotted individually. As a matter of fact, any growing organism is
an open dynamical system getting resources in an exchange with the
surrounding environment (e.g. metabolic exchanges in the case of
biological systems), and only unbounded resources and no spatial
limitations could allow for indefinite growth.

All the systems under investigation, however, are made up of a large
number of individuals (cells for biological systems, atoms or
nucleons for solid state systems or stars, and so on), and an
effective description requires selecting the right set of variables
to represent a specific phenomenon on a chosen scale. For example,
in growing cancers the existence of a multi-scale structure is well
established and this implies a specific approach for each given
scale \cite{Bellomo1,Bellomo2,Lowengrub}. Accordingly, the
scientific investigations include statistical mechanics methods
\cite{Bellomo1,Bellomo2,Drasdo,Alekseev,West,Riffi}, entropic
techniques \cite{Riffi,Yamano,Wrycza}, and stochastic models
\cite{Lande,Gutierrez,Schurz,Skiadas,Khodabin}.

Our attention will be mainly focused on the stochastic models, and
in particular on the logistic and $\theta$-logistic instances, the
Gompertz stochastic model being already rather well established: its
distributions are indeed log-normal and it has been shown that its
macroscopic evolution is properly described by the median of the
process \cite{noi}. The same can not be said, instead, for the
logistic and $\theta$-logistic models, whose solution procedures are
rather more tangled. The key point is that the logistic and
$\theta$-logistic solutions are expressed in terms of exponential
functionals of Brownian motion, convoluted processes of relevant
interest in the financial context \cite{Yor1,Yor2, Yor3}. Exploiting
however their explicit distribution available in the literature
\cite{Yor2}, we are able to provide a closed form for the
distributions at one time of the logistic and $\theta$-logistic
stochastic processes, and the exact expressions of their associated
moments. We provide also the time plots of the sample trajectories,
and a few numerical evaluations of the exact formulas for the most
relevant moments (expectation and variance) to explore their
behavior and their changes at different levels of randomness.

These results are preceded by an analysis of the logistic,
$\theta$-logistic and Gompertz equations in their noiseless,
deterministic layout, with the aim of getting first a perspicuous
and unified interpretation of their structure, and then a more
definite identification of the underlying hypotheses leading to the
macroscopic evolutions. After a look to the form of the equations
with a focus on the important role of time scales, we start again
from the very beginning, i.e. from the task of describing how the
average growth of a system, made up by many individuals, leads to
the macroscopic laws. We show that this result can be deduced from
rather simplified -- but working -- assumptions, with macroscopic
laws connecting percentage increments, and then realizing a
self-controlled evolution. Within this framework we recognize a
$\theta$-hierarchy in dissipating resources, and we also suggest a
unifying procedure accounting for the emergence of the -- seemingly
eccentric -- Gompertz term, by providing a more defined physical
meaning to a known mathematical approach, and by including in so
doing the Gompertz growth in the $\theta$-logistic frame as a
limiting case.

The paper is organized as follows: in the Section~\ref{PDE} we
present the preliminary analysis of the deterministic logistic and
$\theta$-logistic equations in a unified context, with the inclusion
in the same framework of the Gompertz model as a limiting case. The
next Section~\ref{SGM} contains our main results with respect to the
stochastic implementations of the logistic and $\theta$-logistic
models. Here, after summarizing the state of the art including the
explicit stationary distributions and the path-wise solutions of the
stochastic differential equations, we show, by exploiting a few
trajectories simulations and some numerical computation, the strong
impact of the noise intensity on both the process predictability and
its asymptotic expectation. After that we also provide the exact
expressions (in integral form) of the distributions and moments of
the stochastic logistic and $\theta$-logistic processes, along with
some numerical plot of the most important issues (mean and variance)
in the logistic instance, and a concise examination of them.
Discussion and conclusions finally follow in the
Section~\ref{concl}.

\section{Deterministic growth models}
\label{PDE}

\subsection{An overview of known results}
\label{SRE}

In this section we will briefly summarize the main features of the
logistic and Gompertz equations, and we will find out their general
structure in what we regard as their most revealing setting, a
formulation that will provides a hint for later developments. At the
same time we will also put in evidence the important role played by
the time scales. In our models the main variable will the
macroscopic size of the system $n(t)$, namely the (dimensionless)
number of elementary components (e.g. the cells in a biological
systems) at the instant $t$. The $\theta$-logistic equation then
usually takes the form
 \be
\frac{d\,n(t)}{dt} = \omega_e n(t) - \omega_f n^{\theta + 1} (t)
\label{Logeq1}
 \ee
(the \emph{simple logistic} is recovered for $\theta = 1$), while
the Gompertz equation reads
 \be
\frac{d\,n(t)}{dt} = \omega_e n(t) - \omega_f n(t) \, \ln \, n(t)
\label{Geq1}
 \ee
where the constants $\omega_e=\,^1/_{\tau_e}$ and
$\omega_f=\,^1/_{\tau_f}$ are the reciprocal of the characteristic
times $\tau_e$ and $\tau_f$. The $\theta$-logistic equation can also
be recast in the form
 \be \frac{d n (t)}{dt} = \omega_e n(t) \left[1 - \left(\frac{n (t)}{K}\right)^\theta\right], \qquad K =
\left(\frac{\tau_f}{\tau_e}\right)^{\frac{1}{\theta}}=\left(\frac{\omega_e}{\omega_f}\right)^{\frac{1}{\theta}}
\label{Logeq2}
 \ee
while in the Gompertz case we have
 \be \frac{d n (t)}{dt} =
\omega_e n (t) \left(1 - \frac{\ln{n (t)}}{\ln{K}}\right), \qquad K
= e^{\frac{\tau_f}{\tau_e}}= e^{\frac{\omega_e}{\omega_f}}
\label{Geq2}
 \ee
that for later convenience can also be written as
 \be \frac{d \ln{n
(t)}}{dt} = - \omega_f \, \ln\frac{n (t)}{ K} \label{Geq3}
 \ee
The quantity $K$ in the previous equations is the asymptotic value
of $n (t)$ when $t \rightarrow \infty$, i.e. the value of $n$ that
sets its derivative to zero, and that is also known as
\emph{carrying capacity}. It is known that the solutions of our
equations for $n(0)=n_0$ respectively are (see for example
\cite{muller,Salisbury,Khodabin})
\begin{align}
 & n (t) = \frac{K\,n_0}{n_0+(K-n_0)e^{-\omega_et}}&\hbox{(simple logistic)} \label{sollogeq} \\
 & n (t) = \frac{K\,n_0}{\sqrt[\theta]{n_0^\theta+(K^\theta-n_0^\theta)e^{-\theta\,\omega_et}}}&\hbox{($\theta$-logistic)} \label{soltlogeq} \\
 &  n (t) = K\exp\{\alpha_0 \, e^{- \omega_f t}\} \qquad\quad \alpha_0 = \ln (n_0/K)&\hbox{(Gompertz)} \label{solGeq}
\end{align}
Looking back now at the equations\refeq{Logeq2} and\refeq{Geq2}, we
see that they are all of the general form
 \be \frac{d n (t)}{dt} =\omega_en(t) \big[1 - h (n (t))\big] \label{prodform}
 \ee
where $0 < h (n (t)) < 1$, and therefore also $0<1-h(n (t))<1$,
because we always have $n(t)<K$ if -- as it is realistic in our
investigation -- we take $n_0<K$. The second member in the equations
is a product of two terms: the first term, that by himself would
produce an exponential explosion $n_0\,e^{\,\omega_et}$, is
corrected by the second one (a negative feedback, usually known as
\emph{individual growth rate}): it is this counteraction that drives
the system toward its finite asymptotic size. Remark that,
accordingly, one can assume almost vanishing values of $h(n(t))$ at
the early stage of the evolution, the region of time where
Malthusian growth dominates, while the value $1$ is asymptotically
approached for $t \rightarrow \infty$, when the number attains its
maximum value and stops growing.

As for the two characteristic times, it is apparent that $\tau_e$ is
the time scale of the purely exponential growth, while, as emerges
from\refeq{Logeq1} and\refeq{Geq1}, $\tau_f$ characterizes the
\emph{strength} or \emph{speed} of the correcting term. Obviously it
will be $\tau_f > \tau_e$, and usually also $\tau_f \gg \tau_e$. The
carrying capacity emerges from the competition between the
correction and exponential trends, and it is in fact connected with
their ratio: the slower the action of the feedback w.r.t. the
explosion, the larger the carrying capacity. In the Gompertz case
the carrying capacity is the exponential of the said ratio. Since
moreover the whole growth is controlled by the individual growth
rate, the braking mechanism must be linked to the decrease of
resources available for an elementary component of the system.

Before concluding the section, it is useful for later convenience to
introduce a rescaled variable $x (\tau)=x(\omega_et) = n (t)/K$ and
a rescaled time $\tau = \omega_e t$ so that the form of the logistic
and $\theta$-logistic equations respectively become
 \be
\dot{x}(\tau) = x(\tau) \, \big(1 - x (\tau)\big)
\qquad\qquad\dot{x}(\tau) = x(\tau) \, \big(1 - x^\theta
(\tau)\big)\label{RLE}
 \ee
while the corresponding solutions with $x_0=n_0/K$ are
 \be
x(\tau) = \frac{x_0}{x_0 + (1 - x_0)\, e^{- \tau}}  \qquad\qquad
x(\tau)=\left(\frac{x_{0}^{\theta}}{x_{0}^{\theta} + (1 -
x_{0}^{\theta})\, e^{- \theta \, \tau}}\right)^{1/\,\theta}
\label{RSLE}
 \ee

\subsection{Merging the equations}

\subsubsection{General principles of a unified model}

The nonlinear term $h (t)$ in\refeq{prodform} is usually chosen by
resorting to phenomenological criteria depending on the specific
system to be described, or it emerges -- again phenomenologically --
by coupling differential equations as happens, for example, for the
logistic case in the epidemiological context. We propose instead to
get a somewhat more perspicuous description by deriving it from
suitable, albeit still phenomenological, general assumptions. To
this end we will \emph{reboot} our procedure starting again from the
beginning, i.e. from the generally recognized main goal of a
population dynamics inquiry: taken an evolving natural system
consisting, at a given time, of a large number of individuals
components, address the problem of forecasting the growth of this
number at later times. The realistic details of this evolutions
could in fact be rather intricate, and therefore a macroscopic
dynamics should emerge by retrieving suitable averaged quantities
from a fully probabilistic setting. Of course this would require a
very accurate description at a microscopic scale, namely an outright
introduction of stochastic models (see the subsequent
Section~\ref{SGM}). However, a preliminary intermediate approach can
help to shed some light on the whole of these phenomena, and we will
go on here to show that such a kind of approach is possible and
instrumental, in a way reminiscent of what happens to similar
simplified models introduced in very different contexts.

Denoting with $n(t)$ the average number of the elementary components
of our system at the generic instant $t$, the main point is to
compute its increment $\Delta n(t) \doteq n(t+\Delta t)-n(t)$ at a
subsequent time $t+\Delta t$. Here
$\Delta n(t)$ will be supposed to result from the accumulation of
many microscopic increments produced by the possible occurrence of
random events (the birth or death of one individual, one mitosis,
and so on) between $t$ and $t + \Delta t$: at this stage of the
inquiry, however, we will keep this underlying microscopic
\emph{probabilistic} mechanism only in the background. Without yet
assuming a fully stochastic model, indeed, we will only surmise the
existence of this random underworld as a background justification of
our coarse grained deterministic equations. We will moreover assume
the following, simplified hypotheses:
\begin{enumerate}
    \item At each instant, the system can rely
on a finite and fixed (mean) amount of \emph{resources} that we will
(conventionally) denote $E_{\,T}$. The specific nature of these
resources, which can have different origins, is not relevant in our
scheme, because eventually all the quantities will be translated in
terms of number of components
    \item Within the system the individuals exploit
these resources both to \emph{survive}  and to \emph{grow}, but
survival \emph{takes precedence} in the sense that, at each stage,
the resources available for growth are what is left of $E_{\,T}$
once the resources for survival have been taken out. Furthermore, at
each step every individual needs on average a quantity $\epsilon_s$
of resources to survive
    \item Growth stops when the total amount of resources $E_{\,T}$
is only sufficient to the survival of all the individuals: in that
case the population achieves its maximum, finite dimension $K$
a.k.a. \emph{carrying capacity}
    \item\label{probhp} There is a constant, \emph{average rate of
increment per unit time} $\omega_e=\tau_e^{-1}$ of the number of
individuals, so that the average rate of increase in $dt$ will be
$\omega_e\,dt$. In the literature $\omega_e$ is often called
\emph{probability per unit time} and has been already introduced in
very different contexts as, for example, in the Drude simplified
model of conduction \cite{Ashcroft}
\end{enumerate}

Before further developing our model from the previous assumptions,
we consider first an ideal case to provide some suggestions for the
more realistic ones. We will suppose then that there are no
limitations to the available resources ($E_{\,T} = \infty$) and to
the available space. In this case, whatever the need for survival
resources, at any instant the availability of growth resources would
be boundless, and thus the population increment would be obtained by
simply applying the average rate of increase to the whole number
$n(t)$
\begin{equation}\label{Malthusian}
    d n (t) = \omega_e \, n(t)\,dt
\end{equation}
with a resulting Malthusian explosion $n(t) = n_0 \,
e^{\,\omega_et}$. Here of course $n_0$ denotes the system size at
time zero. The previous relation can however be also written as
\begin{equation*}
    \frac{d n (t)}{n (t)} = \omega_e\,dt \times 1
\end{equation*}
On the l.h.s.\ we find the (infinitesimal) percentage increment of
the number, while from the r.h.s.\ we see that this increment
results from the product of the average rate of increment in $dt$
and $1$. Being in our case the available resources not bounded, the
factor $1$ can be simply interpreted as the fraction of resources
available for growth at any instant. On the basis of this
consideration we are led then to propose the following principle:
\begin{quote}
\emph{A growth equation is obtained by imposing that the percentage
increment of a population in a small time interval $dt$ is equal to
the product between  the average rate of increment in the same time
interval, and the percentage of resources (w.r.t.\ the total ones)
that is left available after the survival resources have been used}
\end{quote}
We will see soon that this latter percentage depends only on the
population size.

Going now to more realistic instances, we start from the simplest
case by supposing that \emph{at each instant the resources are
evenly distributed among all the $n(t)$ individuals}. Being
$\epsilon_s$ the mean amount of resources exploited by an individual
to survive, in our approximation we first of all have
\begin{equation*}
   E_{\,T} = \epsilon_s \, K
\end{equation*}
Then, according to our hypotheses, if $n(t) < K$ is the number of
individuals at the instant $t$, the resources exploited for survival
at that instant are $\emph{E}_{s} (t) = \epsilon_s \,n(t) <
E_{\,T}$, and those available for growth are $\emph{E}_g (t) =
E_{\,T}- \emph{E}_{s} (t) = \epsilon_s \, (K - n(t))$ so that
\begin{equation}\label{PG}
    \frac{dn (t)}{n(t)} = \omega_e\,dt
\frac{\emph{E}_g (t)}{E_{\,T}} = \omega_e\,dt \frac{K - n(t)}{K} =
\omega_e\,dt \left(1 - \frac{n(t)}{K}\right)
\end{equation}
and finally in terms of the reduced number and time
\begin{equation}\label{LogisticAgain}
    \frac{d x
(\tau)}{x (\tau)} = d\tau \, (1 - x (\tau))
\end{equation}
that can be easily rearranged into the simple logistic
equation\refeq{RLE} ($\theta = 1$). The result\refeq{Logeq1} can
then be quickly retrieved by reintroducing the variable $n(t)$ and
the characteristic time $\tau_e$, and defining the time $\tau_f =
\tau_e \, K$.

On the other hand -- according to whether the system has a coherent
character, with consequent collective and synergistic behaviors, or,
on the contrary, it displays inefficiencies  and
\emph{non-collaborating} elementary components -- resource scalings
different from the linear one are allowed. A generalized scaling
$E_{\,T} = (\epsilon_s K)^\theta$ and $\emph{E}_s (t) = \epsilon_s
n^\theta(t)$ can thus be introduced, giving rise to the
$\theta$-logistic equation
\begin{equation}\label{ThetaCorrPerc}
    \frac{d x (\tau)}{x (\tau)} = d\tau
\, (1 - x^\theta (\tau))
\end{equation}
In this formulation, however, the Gompertz model still seems to
stand apart: would it be possible to recover even this equation
within the framework of the previous scheme? In the next section we
will provide a path to a positive answer.

\subsubsection{Retrieving the Gompertz equation}

To explain in the above context the eccentric logarithmic term of
the Gompertz model, we must at once recognize that we can no longer
start from some kind of proportionality between the percentage
increase of $n(t)$ and the time interval $\Delta t$. We will instead
suppose more in general for the reduced quantities
\begin{equation}\label{Generalized}
    \frac{\Delta x (\tau)}{x (\tau)} = w (x (\tau), \Delta \tau)
\end{equation}
where $w (x (\tau), \Delta \tau)$ is a function still to be
determined. To this purpose we preliminarily remark that, to be
consistent, the procedure we will establish must anyway lead to a
final result that fulfills some obvious constraints:
\begin{itemize}
    \item $w (x (\tau), \Delta \tau$) must become \emph{small} for large times,
and must \emph{approach} $1$ for small times
    \item $w (x (\tau), \Delta \tau$) must go to zero with
$\Delta \tau$ as a continuity requirement
\end{itemize}
We also expect moreover that, at the end of our procedure, at the
r.h.s.\ of the equation we will find again the product of an
infinitesimal probability times a percentage term constraining the
growth.

We go on now by assuming that $w(x (\tau), \Delta \tau)$ generalizes
the $\theta$-logistic term with the \emph{anomalous scaling} $\theta
(\Delta \tau) = \omega_f\Delta \tau + o (\Delta \tau)$, where
$\tau_f=\omega_f^{-1}$ is the characteristic time-scale. We
therefore take the function
\begin{equation}\label{anscal1}
    w(x (\tau),
\Delta \tau) = 1 - x (\tau)^{\omega_f\Delta \tau + o (\Delta \tau)}
\end{equation}
which apparently fulfills the required constraints: since indeed $K$
is the maximum asymptotic value of $n(t)$, for $t \rightarrow
\infty$ we find $x (\tau) \rightarrow 1$ and the increment of the
number (i.e.\ the correcting term) tends to become small, while in a
very early stage of evolution $x (\tau) \ll 1$ and $w \approx 1$.
The requirement $w(x (\tau), \Delta \tau) \approx 0$ when $\Delta
\tau \approx 0$, is clearly fulfilled as well. We can then take
advantage of a power expansion to write
\begin{equation}
w(x (\tau), \Delta \tau) = 1 - e^{\left(\omega_f\Delta \tau + o
(\Delta \tau)\right) \, \ln \, x (\tau)}=  1 - \big(1 +
\omega_f\Delta \tau \, \, \ln \, x (\tau)\big) + o (\Delta \tau)
\label{equation}
\end{equation}
finding first
\begin{equation}\label{anscalmacr}
    w(x (\tau), d \tau) )=-\omega_f\ln x(\tau)\,d\tau
\end{equation}
and then finally the Gompertz equation\refeq{Geq3} for the reduced
variables
 \be \frac{dx (\tau)}{d\tau} = - \omega_f \,
x (\tau) \, \ln \, x (\tau), \label{gomp}
 \ee
If we remember that $x (\tau) = n(t)/K$, and $\tau_e \doteq (\ln \,
K)^{- 1} \tau_f$, we can also retrace the factorized form
of\refeq{Geq2} as a product of the probability per unit time and a
reduced percentage of available resources. This concludes the
retrieval of the Gompertz model within the framework of our general
scheme.

Remark that the Gompertz growth is obtained when $\theta \rightarrow
0$ in a suitable sense, justifying in this way its \emph{maximally
coherent} character. Moreover, some physical sense can be ascribed
to the the well known mathematical result $1 - x^\theta= -\theta\ln
x+o(\theta)$ when $\theta \rightarrow 0$ often recalled in the
literature when the Gompertz model is investigated: the meaning
indeed is that scaling in the Gompertz growth depends on the
microscopic scales (times) of the system. In turn this fact can
clarify once again the origin of the extremely coherent character of
Gompertz evolution, because the \emph{cooperation level} extends on
the microscopic domain.

\section{Stochastic growth models}
\label{SGM}

We will now discuss a few questions arising from the introduction of
fluctuations and leading to stochastic growth models. Here, the
reduced number $x (\tau)$ will be promoted to a full-fledged
stochastic process $X(\tau)$ in the reduced, dimensionless time
$\tau = \omega_e \, t$, but since from now on there will be no risk
of ambiguity we will revert in the following to the simpler notation
$X(t)$ where it will be always understood that $t$ is the
dimensionless time.

In our scheme it will be rather natural to take fluctuations on the
fraction
\begin{equation*}
    Q_g=\frac{E_g}{E_T}=\frac{E_T-E_s}{E_T}
\end{equation*}
of the resources available for the growth. Considering indeed the
general $\theta$-logistic case and following an usual
procedure~\cite{Khodabin}, we will simply add to $Q_g$ a \emph{white
noise} $\dot{W}(t)$ (namely a process such that $\EXP{\dot{W}(t)}=
0, \; \EXP{\dot{W}(t)\dot{W}(s)} = 2D\, \delta (t - s)$, where $D$
is a constant diffusion coefficient and $\EXP{\,\cdot\,}$ denotes
the expectation) and therefore\refeq{ThetaCorrPerc} will become
\begin{equation}\label{StochImpl}
    \frac{d X(t)}{X(t)} = \big(Q_g + \dot{W}(t)\big)\, dt =
\big[X(t) (1 - X^{\theta} (t)) + \dot{W}(t)\big]\, dt
\end{equation}
giving rise finally to the stochastic differential equation (SDE)
\begin{equation}\label{StochImplDef}
    d X(t) = X(t) \big(1 - X^{\theta} (t)\big)\, dt + X(t)\, dW(t)
\end{equation}
where we exploited the well known fact that the white noise
$\dot{W}(t)$ is the (distributional) derivative of a Wiener process
$W (t)\sim\mathfrak{N}(0,\,2Dt)$ in the sense that $\dot{W}(t)\,dt$
is in fact the increment $dW(t)$ where $\EXP{dW(t)}=0$ and
$\EXP{dW(t)dW(s)}= 2D\, \delta (t - s)\,dtds$. Remark that with this
procedure, whatever the growth law considered, the stochastic term
is always given by $X \, dW$: this term is widely adopted in the
literature about the logistic and $\theta$-logistic cases, although
multiplicative noises, or even more complex additive stochastic
terms, have been introduced both in discrete and continuous time
versions
\cite{Schurz,Skiadas,Khodabin,Bartlett,Ovaskainen2,Pasquali,Nasell1,Nasell2,Nasell3,Ramasubramanian,Tenkes}.
In the Gompertz instance, adding this noise term directly leads to
the a geometric Wiener process and, as pointed out in the
introduction, in this case all the aspects of the model, and its
connection with the macroscopic equation, are completely defined.
For the stochastic logistic and $\theta$-logistic models instead
only a few aspects have been completely elaborated, while others,
and very important too, still are not. In the following, we first
summarize the results already obtained in the literature, and then
we discuss our main new results.
\begin{figure}
\begin{center}
\includegraphics*[width=15cm]{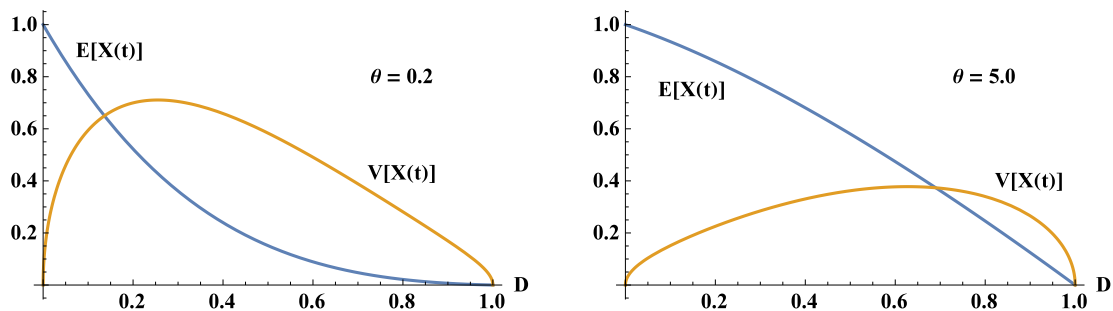}
\caption{Expectations and variances of stationary $\theta$-logistic
processes as a function of $D$ and for several values of
$\theta$}\label{statmom2}
\end{center}
\end{figure}

\subsection{A few preliminary results about the logistic models}
\label{KRSGM}

Many aspects of the logistic and $\theta$-logistic stochastic models
have been already systematically discussed (see for
instance~\cite{ReportNoi}): we will recall here just a few relevant
results useful in the following sections. First, the
\emph{stationary distributions} have been computed and their
stability has been studied too \cite{Pasquali}; also
quasi-stationary distributions have been investigated in the
discrete case \cite{Ovaskainen,Nasell1,Nasell2}. The stationary
distribution for the stochastic $\theta$-logistic equation is the
generalized gamma law
$\mathfrak{G}_\theta\left(\frac{1-D}{D},\frac{1}{(\theta
D)^{1/\theta}}\right)$ with \emph{pdf}
\begin{equation}\label{SDSTLE}
    f_s(x) = \frac{\theta\, x^{\frac{1-D}{D}-1}e^{-\frac{x^\theta}{\theta D}}}{(\theta D)^{\frac{1-D}{\theta D}}\Gamma\left(\frac{1-D}{\theta D}\right)}
\end{equation}
provided that $D < 1$. This last condition ensures normalization,
and defines the region of stability of the system. The simple
logistic case is obtained by choosing $\theta = 1$. (for
computational details, see also \cite{ReportNoi}). It is also easy
to see then that the moments in the stationary
distribution\refeq{SDSTLE} are
\begin{equation}\label{statmom}
    \EXP{X^k(t)}=(\theta D)^{\frac{k}{\theta}}\frac{\Gamma\left(\frac{1+(k-1)D}{\theta D}\right)}{\Gamma\left(\frac{1-D}{\theta D}\right)}
\end{equation}
and in particular for the simple logistic ($\theta=1$) we have
$\EXP{X(t)}=1-D$ and $\bm V[X(t)]=D(1-D)$. These simple results (and
their generalizations for the $\theta$-logistic cases shown in the
Figure \ref{statmom2}) suggest that the asymptotic (ergodic)
stationary level of a random logistic is in average suppressed by
high noise intensity ($D$ near to $1$). In other words, the noise
acts as an effective disruption on the logistic growth: a relevant
point that will be resumed later.

Even the \emph{path-wise solutions} of the processes are explicitly
known \cite{Skiadas,ReportNoi}. If indeed we define the following
Wiener process with constant drift
\begin{figure}
\begin{center}
\includegraphics*[width=8cm]{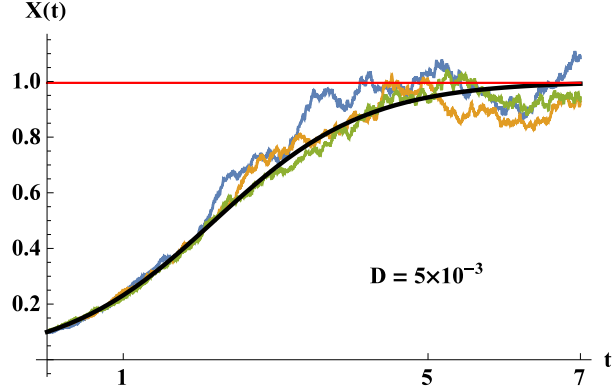}
\caption{Sample paths of a simple logistic $X(t)$ with $D=0.005$.
The horizontal red line represents the asymptotic, stationary
expectation}\label{path1}
\end{center}
\end{figure}
\begin{equation}\label{Z}
    Z(t)=(1-D)t+W(t)\sim\mathfrak{N}\big((1-D)t,\,2Dt\big)
\end{equation}
it is possible to show that the solution of the $\theta$-logistic
SDE\refeq{StochImplDef} with initial condition $X(0)=X_0,\;\bm
P\hbox{-\emph{a.s.}}\;$ is
\begin{equation}\label{SolProcTL}
    X(t)=\left(\frac{X_0^\theta\, e^{\,\theta Z(t)}}{1+\theta
    X_0^\theta\int_0^te^{\,\theta Z(u)}du}\right)^{1/\theta}
\end{equation}
that is correctly brought back to the noiseless, deterministic
solution\refeq{RSLE} by switching off the noise ($D=0$ and
$W(t)=0,\;\bm P\hbox{-\emph{a.s.}}$, namely $Z(t)=t$) and by taking
a degenerate initial condition $X_0=x_0,\;\bm P\hbox{-\emph{a.s.}}$
The solution of the simple logistic SDE\refeq{StochImplDef} with
$\theta=1$ finally is
\begin{equation}\label{SolProcL}
    X(t)=\frac{X_0\, e^{\,Z(t)}}{1+
    X_0\int_0^te^{\,Z(u)}du}
\end{equation}

\subsection{Sample paths, distributions and moments}\label{pdm}

Despite the expressions\refeq{SolProcTL} and\refeq{SolProcL} being
fully explicit, to compute the (non-stationary) expectation
$\EXP{X(t)}$ and the higher moments $\EXP{X^k(t)}$ is not at all a
simple task, and since not even a perturbative approach in terms of
small noisy disturbances seems to be available \cite{Gardiner}, the
fully non-perturbative tools will be in fact required. Looking at
the expressions\refeq{SolProcTL} and\refeq{SolProcL} we see on the
other hand that the integrals in the denominators (the terms hardest
to crack) are indeed processes usually called \emph{exponential
functionals of Brownian motion} (EFBM) of the type
\begin{figure}
\begin{center}
\includegraphics*[width=8cm]{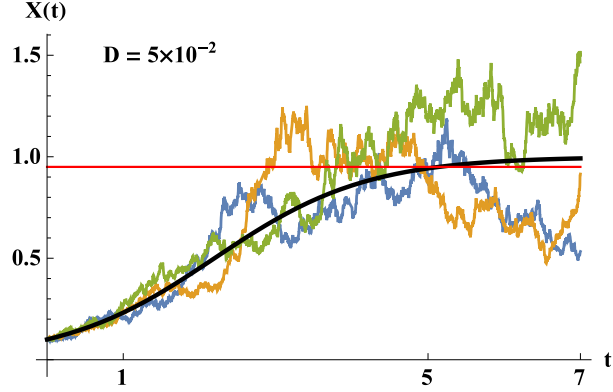}
\caption{Sample paths of a simple logistic $X(t)$ with $D=0.05$. The
horizontal red line represents the asymptotic, stationary
expectation}\label{path2}
\end{center}
\end{figure}
\begin{equation}\label{EFBMDef}
    \int_{0}^{t} e^{\,aW (u) + b\,u}du
\end{equation}
that have been extensively studied in the financial context
\cite{Yor1,Yor2, Yor3}. Remark that since the Wiener process is
Gaussian we have $W(t)\sim\mathfrak{N}(0,2Dt)$, and therefore it is
also $\theta
Z(t)\sim\mathfrak{N}\big(\theta(1-D)t,2\theta^2Dt\big)$. As a
consequence the integrand of our EFBM is log-normal $e^{\,\theta
Z(t)}\sim\mathfrak{lnN}\big(\theta(1-D)t,2\theta^2Dt\big)$ and the
following expectations are easily calculated
\begin{equation}\label{efbmEXP}
    \EXP{e^{\,\theta Z(t)}}=e^{\theta\,[1+(\theta-1)D]\,t}\qquad\quad\EXP{\int_0^te^{\,\theta Z(u)}du}=\frac{e^{\theta\,[1+(\theta-1)D]\,t}-1}{\theta\,[1+(\theta-1)D]}
\end{equation}
Many other results about these EFBM are collected in the literature
\cite{Yor1,Yor2, Yor3}, but their exact distributions are rather
convoluted, and on the other hand the determination of the moments
of\refeq{SolProcTL} and\refeq{SolProcL} requires precisely the
utilization of these tangled joint distributions of $Z(t)$ with its
corresponding EFBM. In the following we will therefore provide a few
exact formulas for the probability density functions (\emph{pdf})
and the moments of our process $X(t)$, along with some numerical
estimate of the values of these moments.
\begin{figure}
\begin{center}
\includegraphics*[width=8cm]{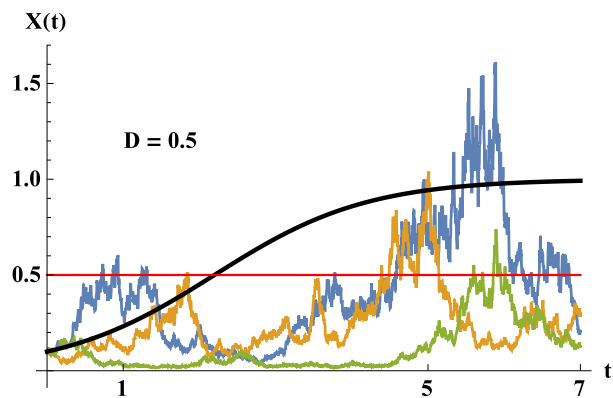}
\caption{Sample paths of a simple logistic $X(t)$ with $D=0.5$. The
horizontal red line represents the asymptotic, stationary
expectation}\label{path3}
\end{center}
\end{figure}

\subsubsection{Trajectories simulations}

\begin{figure}
\begin{center}
\includegraphics*[width=8cm]{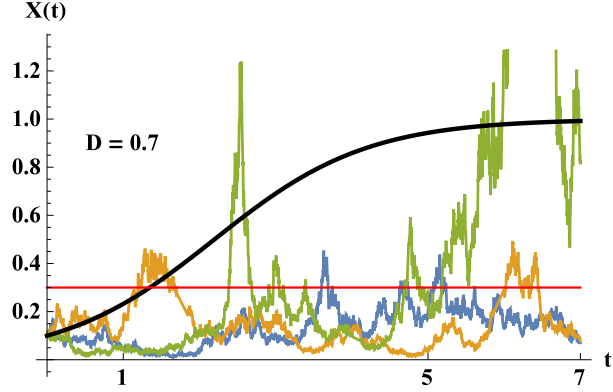}
\caption{Sample paths of a simple logistic $X(t)$ with $D=0.7$. The
horizontal red line represents the asymptotic, stationary
expectation}\label{path4}
\end{center}
\end{figure}

\begin{figure}
\begin{center}
\includegraphics*[width=8cm]{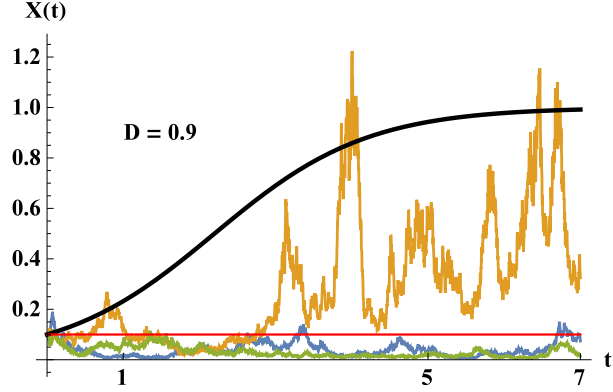}
\caption{Sample paths of a simple logistic $X(t)$ with $D=0.9$. The
horizontal red line represents the asymptotic, stationary
expectation}\label{path5}
\end{center}
\end{figure}

\begin{figure}
\begin{center}
\includegraphics*[width=8cm]{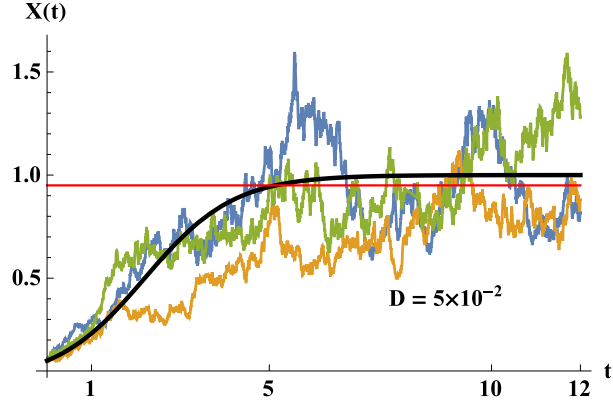}
\caption{Ergodic relaxation in time toward stationary fluctuations
around the asymptotic expectation (red line)}\label{path6}
\end{center}
\end{figure}

We will stop first, however, to present a few numerical simulations
of the sample trajectories of the process $X(t)$ confining ourselves
for clarity to the simple logistic case\refeq{SolProcL} with
$\theta=1$. We will progressively turn the noise on by increasing
the diffusion coefficient $D$, and we will compare the random paths
of the process with both its deterministic behavior (the smooth,
monotonic black curve) and its asymptotic, stationary expectation
(the horizontal, red line). It is apparent then from the first pair
of plots in the Figures \ref{path1} and \ref{path2} that for a
reasonably low level of noise (here $D$ is either $0.005$ or $0.05$)
the random paths fluctuate close to the deterministic curve, and
then asymptotically stabilize around their ergodic expectation.
Moreover the stationary variance grows with $D$. When on the other
hand the value of the diffusion coefficient increases toward $0.5$
or $0.7$ as in the Figures \ref{path3} and \ref{path4} the behavior
of the trajectories begins to be much more irregular with spikes and
flat spots surrounding a decreasing asymptotic expectation. If
finally $D$ approaches the value $1$ (we remember that in order to
find a possible stationary solution we must suppose $D<1$) the
random samples in the Figure \ref{path5} become quite unpredictable
with paths that mostly never take off, while a few other
trajectories briefly explode to larger values: asymptotically
however the paths crash near to zero. Finally in the Figure
\ref{path6} the ergodic relaxation toward the stationary fluctuation
(the variability of the paths looks indeed to be stabilized) is
apparent when we consider a somewhat longer time span. As a matter
of fact our pictures display just a few examples, but the general
conduct of the trajectories seems in fact to be already well
sketched out and is in perfect agreement with the remarks about the
stationary solutions put forward in the Section \ref{KRSGM}.

\subsubsection{A reformulation in terms of the standard Brownian motion}

In order to be able to take advantage more easily of the results
existing in the literature we will first convert our previous
formulas into a slightly different, customary notation~\cite{Yor2}:
to this purpose we introduce the standard Brownian motion $B_t\sim
\mathfrak{N}(0,\,t)$ and its corresponding EFBM
\begin{align}
    & B_t^{(\nu)}=B_t+\nu t\sim\mathfrak{N}(\nu
    t,\,t)\qquad 2B_t^{(\nu)}=2B_t+2\nu t\sim\mathfrak{N}(2\nu
    t,\,4t)\\
    &A_t^{(\nu)}=\int_0^te^{2B_s^{(\nu)}}ds=\int_0^te^{2(B_s+\nu
    s)}ds\qquad\qquad A_t=A_t^{(0)}\label{yorA}
\end{align}
and then using the self-similarity properties of a Wiener process
\begin{equation*}
    \sqrt{\lambda}\,W(t)=W(\lambda t)\qquad\quad B_s=\frac{W(s)}{\sqrt{2D}}=W\left(\frac{s}{2D}\right)\quad\qquad
    \sqrt{2D}\,B_t=B_{2Dt}=W(t)
\end{equation*}
we can reduce our previous formulas to this new notation. First with
the change of integration variable
\begin{equation*}
    s= \frac{D\theta^2}{2}\,u
\end{equation*}
we have
\begin{equation*}
    \int_0^te^{\,\theta Z(u)}du=\frac{2}{D\theta^2}\int_0^{D\theta^2t/2}e^{\theta
    Z\left(\frac{2s}{D\theta^2}\right)}ds=\frac{2}{D\theta^2}\int_0^{\tau}e^{\theta
    Z\left(\frac{2s}{D\theta^2}\right)}ds\qquad\qquad\tau=\frac{D\theta^2}{2}\,t
\end{equation*}
On the other hand we have
\begin{eqnarray*}
  \theta Z\left(\frac{2s}{D\theta^2}\right) &=& \theta W\left(\frac{2s}{D\theta^2}\right)+\frac{1-D}{D\theta}2s
  =2W\left(\frac{s}{2D}\right)+\frac{1-D}{D\theta}2s \\
   &=&2B_s+\frac{1-D}{D\theta}2s=2(B_s+\nu s)=2B_s^{(\nu)}\qquad\qquad\nu=\frac{1-D}{D\theta}
\end{eqnarray*}
and hence
\begin{equation}\label{}
   \int_0^te^{\,\theta
   Z(u)}du=\frac{2}{D\theta^2}\int_0^{\tau}e^{2B_s^{(\nu)}}ds=\frac{2A_\tau^{(\nu)}}{D\theta^2}\qquad\qquad\tau=\frac{D\theta^2}{2}\,t\qquad\nu=\frac{1-D}{D\theta}
\end{equation}
This puts the denominator of\refeq{SolProcTL} in terms
of\refeq{yorA}. Now we must reduce also the numerator to a function
of the exponential of $B_\tau^{(\nu)}$ with the same $\tau$ and
$\nu$ of $A_\tau^{(\nu)}$. Since we have
\begin{eqnarray*}
  \theta Z(t) &=& \theta W(t)+(1-D)\theta t=2W\left(\frac{\theta^2t}{4}\right)+(1-D)\theta t=2B_{\frac{D\theta^2}{2}\,t}+(1-D)\theta t \\
   &=&2\left(B_\tau+\frac{(1-D)\theta}{2}t\right)=2\left(B_\tau+\nu\tau\right)=2B_\tau^{(\nu)}
\end{eqnarray*}
the formula\refeq{SolProcTL} for the process paths in terms of
$A_\tau^{(\nu)}$ and $B_\tau^{(\nu)}$ finally becomes
\begin{equation}
    X(t) = \left(\frac{x_0^\theta\,e^{\,2B_\tau^{(\nu)}}}{1+\frac{2x_0^\theta}{D\theta}
    A_\tau^{(\nu)}}\right)^{\frac{1}{\theta}}=\left(\frac{D\theta \,x_0^\theta\,e^{\,2B_\tau^{(\nu)}}}{D\theta+2x_0^\theta
    A_\tau^{(\nu)}}\right)^{\frac{1}{\theta}}
    \qquad\quad\tau=\frac{D\theta^2}{2}\,t\quad\nu=\frac{1-D}{D\theta}\label{kmom2}
\end{equation}
where $D>0$, $\tau>0$ and $\nu>-1$. This will give us in the
following the possibility of directly exploiting a few preexisting
results.

\subsubsection{Probability density functions}

We know (see for instance~\cite{Yor2}) that the joint \emph{pdf} of
$A_\tau^{(\nu)},\,B_\tau^{(\nu)}$ in their respective values $a$ and
$b$ is
\begin{eqnarray}
  g(a,b)& = &\frac{e^{\,\nu\, b\,-\frac{\nu^2\tau}{2}-\frac{1+e^{2b}}{2a}}}{a}\,\vartheta\left(\frac{e^b}{a}\,,\tau\right)\nonumber\\
  &=&\frac{e^{-\frac{\nu^2\tau}{2}+\frac{\pi^2}{2\tau}}\,
           e^{\,(\nu+1)\, b\,-\frac{1+e^{2b}}{2a}}}{a^2\sqrt{2\pi^3\tau}}\int_0^\infty
  e^{-\frac{e^b}{a}\cosh s}\sinh s\,\,e^{-\frac{ s^2}{2\tau}}\sin\frac{\pi s}{\tau}\,\,ds\label{ABjoint}\\
  \vartheta(r,v)&=&\frac{r\,e^{\frac{\pi^2}{2v}}}{\sqrt{2\pi^3v}}\int_0^\infty
  e^{-\frac{ s^2}{2v}\,-r\cosh s}\sinh s\,\sin\frac{\pi
  s}{v}\,\,ds\label{theta}
\end{eqnarray}
and therefore, in addition to being able to simulate trajectories,
we are also in a position to calculate both the \emph{pdf} of $X(t)$
and its moments. We see indeed from\refeq{kmom2} that $X(t)$ is a
function of $A_\tau^{(\nu)}$ and $B_\tau^{(\nu)}$, and being
apparently $A_\tau^{(\nu)}\ge0$ it is also easy to realize that
\begin{equation*}
   Y(t)= e^{B_\tau^{(\nu)}}\ge\left(\frac{X(t)}{x_0}\right)^\frac{\theta}{2}
\end{equation*}
We can then first find the joint \emph{pdf} $h(x,y)$ of $X(t)$ and
$Y(t)$ with the following monotone variable transformation
\begin{equation}\label{transf}
    \left\{
      \begin{array}{l}
        x=\left(\frac{D\,\theta \,x_0^\theta\,e^{\,2b}}{D\,\theta+2x_0^\theta a}\right)^{1/\theta}\ge0 \\
        y=e^{\,b}\ge\left(\frac{x}{x_0}\right)^{\theta/2}\ge0
      \end{array}
    \right.
    \qquad\qquad
    \left\{
      \begin{array}{l}
        a=\frac{D\,\theta}{2}\left(\frac{y^2}{x^\theta}-\frac{1}{x_0^\theta}\right)\ge0\\
        b=\ln y
      \end{array}
    \right.
\end{equation}
and afterwards calculate the univariate \emph{pdf} of $X(t)$ by
simple marginalization. The Jacobian of the transformation being
\begin{equation*}
    J=\begin{vmatrix}
        ^{\partial x}\!/\!_{\partial a} & ^{\partial x}\!/\!_{\partial b} \\
        ^{\partial y}\!/\!_{\partial a} & ^{\partial y}\!/\!_{\partial b} \\
      \end{vmatrix}
      =\begin{vmatrix}
         ^{\partial x}\!/\!_{\partial a} & ^{\partial x}\!/\!_{\partial b} \\
         0 & e^{\,b} \\
       \end{vmatrix} =e^{\,b}\frac{\partial x}{\partial
       a}=-\frac{2e^{-b}}{D\theta^2}\left(\frac{D\,\theta \,x_0^\theta\,e^{\,2b}}{D\,\theta+2x_0^\theta a}\right)^{\frac{1+\theta}{\theta}}
       =-\frac{2x^{1+\theta}}{D\theta^2\,y}
\end{equation*}
the new joint \emph{pdf} is
\begin{equation}\label{jointpdf}
    h(x,y)=\frac{g\big(a(x,y),\,b(x,y)\big)}{|J(x,y)|}
\end{equation}
so that from\refeq{ABjoint} with $y\ge(x/x_0)^{\theta/2}$ we have
\begin{eqnarray}
  h(x,y) &=& \frac{e^{-\frac{\nu^2\tau}{2}+\frac{\pi^2}{2\tau}}}{\sqrt{2\pi^3\tau}}\,\frac{2x_0^{2\theta}\,x^{\theta-1}y^{\nu+2}}{D\,(x_0^\theta y^2-x^\theta)^2}
   \,e^{-\frac{x_0^\theta x^\theta(1+y^2)}{D\theta(x_0^\theta
   y^2-x^\theta)}}\\
   &&\qquad\qquad\qquad\times\int_0^\infty ds\,\, e^{-\frac{2\,x_0^\theta x^\theta y}{D\theta(x_0^\theta y^2-x^\theta)}\cosh s}
   e^{-\frac{ s^2}{2\tau}}\sinh s\,\sin\frac{\pi s}{\tau}\nonumber
\end{eqnarray}
and finally, with the further change of variable $u=x_0^\theta
y^2-x^\theta$, the \emph{pdf} of $X(t)$ is
\begin{eqnarray}
  f(x,t) &=& \int_0^\infty h(x,y)\,dy\;=\;\frac{e^{-\frac{\nu^2\tau}{2}+\frac{\pi^2}{2\tau}}}{\sqrt{2\pi^3\tau}}\int_{\left(\frac{x}{x_0}\right)^{\theta/2}}^\infty dy\frac{2x_0^{2\theta}\,x^{\theta-1}y^{\nu+2}}{D\,(x_0^\theta y^2-x^\theta)^2}
    \nonumber\\
   &&\qquad\qquad\qquad\qquad\qquad\times\int_0^\infty ds\,\, e^{-\frac{x_0^\theta x^\theta (1+2y\cosh z+y^2)}{D\theta(x_0^\theta y^2-x^\theta)}}
   e^{-\frac{ s^2}{2\tau}}\sinh s\,\sin\frac{\pi s}{\tau}\nonumber\\
   &=&\frac{x_0^{\frac{(1-\nu)\theta}{2}}e^{-\frac{\nu^2\tau}{2}+\frac{\pi^2}{2\tau}}}{D\sqrt{2\pi^3\tau}}\,x^{\theta-1}
   \int_0^\infty du\frac{(u+x^\theta)^{\frac{\nu+1}{2}}}{u^2}\label{Xpdf}\\
   &&\qquad\qquad\times\int_0^\infty ds\,\, e^{-\frac{x^\theta}{D\,\theta u}\big(x_0^\theta+2x_0^{\theta/2 }\sqrt{u+x^\theta}\cosh s+u+x^\theta\big)}
   e^{-\frac{ s^2}{2\tau}}\sinh s\,\sin\frac{\pi s}{\tau}\nonumber
\end{eqnarray}
In particular, in the case of a simple logistic ($\theta=1$) we have
\begin{eqnarray}
    f(x,t)&=&\frac{x_0^{\frac{1-\nu}{2}}e^{-\frac{\nu^2\tau}{2}+\frac{\pi^2}{2\tau}}}{D\sqrt{2\pi^3\tau}}
   \int_0^\infty du\frac{(u+x)^{\frac{\nu+1}{2}}}{u^2}\\
   &&\qquad\qquad\quad\times\int_0^\infty ds\,\, e^{-\frac{x}{D\,u}\big(x_0+2\sqrt{x_0(u+x)}\cosh s+u+x\big)}
   e^{-\frac{ s^2}{2\tau}}\sinh s\,\sin\frac{\pi s}{\tau}\nonumber
\end{eqnarray}

\subsubsection{Moments of $X(t)$}

The moments of $X(t)$ can now be calculated either directly
form\refeq{kmom2} and\refeq{ABjoint} as
\begin{eqnarray}
  \EXP{X^k(t)} &=& \int_0^\infty da\int_{-\infty}^\infty db\,\left(\frac{D\theta \,x_0^\theta\,e^{\,2b}}{D\theta+2x_0^\theta a}\right)^{\frac{k}{\theta}}  g(a,b)\nonumber \\
   &=&\int_0^\infty da\int_{-\infty}^\infty db\,\left(\frac{D\theta \,x_0^\theta\,e^{\,2b}}{D\theta+2x_0^\theta a}\right)^{\frac{k}{\theta}}
   \frac{e^{-\frac{\nu^2\tau}{2}+\frac{\pi^2}{2\tau}} e^{\,(\nu+1)\,
   b\,-\frac{1+e^{2b}}{2a}}}{a^2\sqrt{2\pi^3\tau}}\label{mom}\\
   &&\qquad\qquad\qquad\qquad\qquad\qquad \times\int_0^\infty e^{-\frac{e^b}{a}\cosh s}e^{-\frac{ s^2}{2\tau}}\sinh s\,\sin\frac{\pi
   s}{\tau}\,\,ds\nonumber
\end{eqnarray}
or from the marginal \emph{pdf}\refeq{Xpdf} of $X(t)$ as
\begin{eqnarray}
  \EXP{X^k(t)} &=& \int_0^\infty x^kf(x,t)\,dx\nonumber\\
   &=&\frac{x_0^{\frac{(1-\nu)\theta}{2}}e^{-\frac{\nu^2\tau}{2}+\frac{\pi^2}{2\tau}}}{D\sqrt{2\pi^3\tau}}
   \int_0^\infty dx\,x^{\theta+k-1}
   \int_0^\infty du\frac{(u+x^\theta)^{\frac{\nu+1}{2}}}{u^2}\label{Xmom}\\
   &&\qquad\quad\times\int_0^\infty ds\,\, e^{-\frac{x^\theta}{D\,\theta u}\big(x_0^\theta+2x_0^{\theta/2 }\sqrt{u+x^\theta}\cosh s+u+x^\theta\big)}
   e^{-\frac{ s^2}{2\tau}}\sinh s\,\sin\frac{\pi s}{\tau}\nonumber
\end{eqnarray}
In particular the first moment (expectation) of the simple logistic
($\theta=1$) in the two formulations is
\begin{figure}
\begin{center}
\includegraphics*[width=10cm]{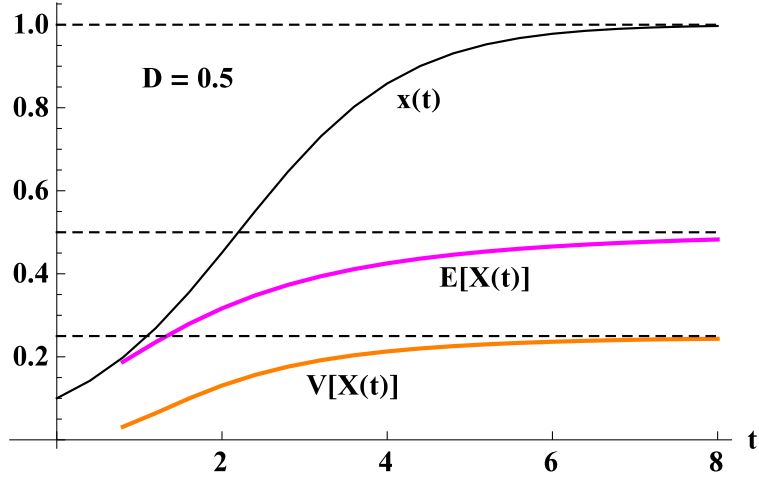}
\caption{Time-dependent behavior of the expectation (magenta) and
variance (orange) of a non stationary, simple logistic process with
$D=\,^1/_2$ and degenerate initial condition $x_0=0.1$ as computed
from\refeq{mom}. The two moments ergodically tend to their
asymptotic stationary values (dashed lines respectively at $0.50$
and $0.25$) and are here compared to the noiseless growth $x(t)$
with the same initial condition}\label{nonstat}
\end{center}
\end{figure}
\begin{eqnarray}
  \EXP{X(t)} &=& \int_0^\infty da\int_{-\infty}^\infty db\,\frac{D
  \,x_0\,e^{\,2b}}{D+2x_0a}\,
   \frac{e^{-\frac{\nu^2\tau}{2}+\frac{\pi^2}{2\tau}} e^{\,(\nu+1)\,
   b\,-\frac{1+e^{2b}}{2a}}}{a^2\sqrt{2\pi^3\tau}}\\
   &&\qquad\qquad\qquad\qquad\qquad\qquad \times\int_0^\infty e^{-\frac{e^b}{a}\cosh s}e^{-\frac{ s^2}{2\tau}}\sinh s\,\sin\frac{\pi
   s}{\tau}\,\,ds\nonumber\\
   &=&\frac{x_0^{\frac{1-\nu}{2}}e^{-\frac{\nu^2\tau}{2}+\frac{\pi^2}{2\tau}}}{D\sqrt{2\pi^3\tau}}
   \int_0^\infty dx\,x\int_0^\infty du\frac{(u+x)^{\frac{\nu+1}{2}}}{u^2}\\
   &&\qquad\qquad\quad\times\int_0^\infty ds\,\, e^{-\frac{x}{D\,u}\big(x_0+2\sqrt{x_0(u+x)}\cosh s+u+x\big)}
   e^{-\frac{ s^2}{2\tau}}\sinh s\,\sin\frac{\pi s}{\tau}\nonumber
\end{eqnarray}
The multiple integrals listed in the present section can not
apparently be performed analytically and should therefore be
computed numerically. This integration is rather tricky due to the
presence of the inner oscillating integral\refeq{theta}. Even with
spartan computational tools however it is possible to check that a
number of available preliminary results are fully consistent with
the previous theoretical forecasts. Taking for instance the
non-stationary simple logistic process\refeq{SolProcL} with
$\theta=1$, $D=\,^1/_2$ and degenerate initial condition $x_0=0.1$,
a numerical evaluation of the first two moments in a time interval
from $0.8$ to $8.0$ leads to the time depending behavior of
expectation and variance displayed in the Figure~\ref{nonstat}. By
ideally extrapolating the plots to $t=0$ it is easy to see then that
$\EXP{X(t)}$ and $\bm V[X(t)]$ steadily and monotonically grow from
their initial values (respectively $0.1$ and $0.0$) toward their
asymptotic, stationary values $0.50$ and $0.25$, so that in
particular the asymptotic average level of the process stays well
below the deterministic curve $x(t)$ of\refeq{RSLE} as already
anticipated in the Section~\ref{KRSGM}. The consistency of these
simple result hints therefore to the fact that the exact, closed
formulas presented in the present section can be now confidently
adopted for every calculation regarding the non stationary logistic
and $\theta$ logistic processes if one can master a few routine
difficulties in the integration procedure.

\subsubsection{The logistic transition {pdf}}

Also the computation of the logistic transition \emph{pdf}'s is a
demanding task that stimulated  numerical investigations
too~\cite{Khodabin,Tenkes}. By exploiting a further general formula
known in the literature \cite{Gihman} we will provide here another
closed expressions for the transition \emph{pdf}'s of the
SDE\refeq{StochImplDef} whose finalization however again requires
the calculation of some particular expectation: for more details
about the derivation procedure see \cite{ReportNoi}. For the simple
logistic and the $\theta$-logistic processes we indeed respectively
have
\begin{eqnarray}
    f (x, t|y, s)& =&g(x,t;y,s) \,
\EXP{G (x, t; y, s)}\label{TPDFSLE} \\
    f_\theta (x, t|y, s)&=& g_\theta(x,t;y,s)\, \EXP{G_\theta (x, t;y,s)}\label{TPDFSTLE}
\end{eqnarray}
where, by taking advantage of the following Brownian bridge between
$\overline{W}_{st}(0)=0$ and $\overline{W}_{st}(1)=0$
\begin{equation}\label{bridge}
    \overline{W}_{s t} (r) = W (s + (t - s) r) - [r W (t) + (1 - r) W
    (s)]\qquad\quad0\le r\le1
\end{equation}
we have defined
\begin{align}
   &\qquad\qquad\qquad\qquad g(x,t;y,s)=\frac{e^{- \frac{x - y}{2D} - \frac{1}{4 D (t - s)} \left[(1 - D)
(t - s) - \ln{\frac{x}{y}}\right]^2}}{x \sqrt{4 \pi D (t - s)}}\\
   &\qquad\qquad\qquad\qquad g_\theta(x,t;y,s)=\frac{e^{- \frac{x^\theta - y^\theta}{2 D \theta}
- \frac{1}{4 D (t - s)} \left[(1 - D) (t - s) -
\ln{\frac{x}{y}}\right]^2}}{x
\sqrt{4 \pi D (t - s)}}\\
   &\qquad\qquad G (x, t; y, s) = e^{- \frac{t - s}{4 D} \, H (x,
t; y, s)}\qquad\qquad G_\theta (x, t; y, s) = e^{- \frac{t - s}{4 D}
\, H_\theta (x, t; y, s)} \label{Gtheta}\\
   &H (x, t; y, s) \,=\,
y^2 \int_{0}^{1} dr \left(\frac{x}{y}\right)^{2 r} e^{2
\overline{W}_{s t} (r)} - 2  y \int_{0}^{1} dr
\left(\frac{x}{y}\right)^{r} e^{\overline{W}_{s t} (r)} \label{g}\\
   &H_\theta (x, t; y, s)\, =\, y^{2  \theta} \int_{0}^{1} dr
\left(\frac{x}{y}\right)^{2  \theta  r} e^{2  \theta \overline{W}_{s
t} (r)}\nonumber\\
&\qquad\qquad\qquad\qquad\qquad\qquad - 2  \big[1 + (\theta - 1)
D\big] y^{\theta} \int_{0}^{1} dr \left(\frac{x}{y}\right)^{\theta
r} e^{\theta \overline{W}_{s t} (r)}\label{gtheta}
\end{align}
The expected values contained in the above formulas can again be
computed exactly by following the same steps presented in the
previous sections because apparently they are once more expressed in
terms of particular EFBM's and their distributions can therefore be
traced back to the \emph{pdf}\refeq{ABjoint}. We will neglect
however an explicit calculation for the sake of brevity.

\section{Conclusions and outlook}\label{concl}

In the present paper we presented several exact results referring to
the stochastic logistic and  $\theta$-logistic models. Before
dealing with these random instances, however, we preliminarily
performed a careful analysis of the deterministic, noiseless
logistic and $\theta$-logistic growths, showing that they can be
discussed in an unified context where the dynamics emerges from the
proportionality between the relative increment of the number of
elementary individuals and the percentage of resources exceeding the
needs for the simple subsistence. The parameter $\theta$ is moreover
interpreted as characterizing the level of correlation (classical
coherence) among the individuals present in a system: in particular
the correlation increases as $\theta$ decreases. In this framework,
the Gompertz model -- retrieved when $\theta$ goes to zero in a
suitable sense -- is placed by an anomalous scaling at the top of
the hierarchy as \emph{the more coherent one}.

In the  second part of the article, we went on to  deal with
stochastic logistic and  $\theta$-logistic models. After introducing
the random fluctuations in agreement with our previous principles,
we summarized the known results about the stochastic logistic and
$\theta$-logistic SDE's, i.e. their stationary distributions and
their path-wise solutions. We performed next a few trajectories
simulations whose inspection turns out to be instrumental to show
that -- whereas at a reasonably low level of noise the random paths
fluctuate close to the deterministic curve, and then asymptotically
stabilize around their ergodic expectation -- a sensible increase of
the noise intensity effectively destabilizes the process, making its
behavior on the one hand more and more unpredictable, and on the
other asymptotically vanishing in average as predicted in the
stationary solutions.

We provided next our main results, i.e.  the exact expressions (in
an integral closed form) of the probability distributions and
moments of the stochastic logistic and $\theta$-logistic processes,
deducing -- with a suitable change of variable and a marginalization
-- their probability density functions from the joint distribution
of a Brownian process and its associated EFBM already known in the
literature~\cite{Yor1,Yor2,Yor3}. In the simple logistic case
($\theta=1$) a numerical computation of the time-behavior of
expectation and variance was performed for a given noise intensity,
showing that their values monotonically grow in time, and that they
ergodically tend to their asymptotic, stationary values. In
addition, we also provided a semi-explicit closed form for the
transition \emph{pdf} of the logistic SDE's, from which a fully
explicit expression can be obtained by taking advantage of the same
distributions previously exploited. We preferred however to postpone
this computation to a possible forthcoming publication for the sake
of brevity: we look forward indeed to extend these methods to obtain
further exact or approximate results for other complex stochastic
models describing more specific systems, and to deal with several
unanswered questions.

Among the open problems, in particular, that of finding a suitable
coarse-grained version of the logistic SDE's certainly is
outstanding. We have shown in the previous sections that for $D\to0$
the trajectories and the moments of a $\theta$-logistic process
apparently inch closer and closer to the deterministic behavior of a
noiseless growth. This is a feature that the logistic models share
with the Gompertz one, and of course it is what we were looking for
in a stochastic model correctly generalizing a deterministic one. At
least in the Gompertz case, however, it was proved in a previous
paper~\cite{noi} that there is something more: it is possible indeed
to \emph{coarse-grain} the model SDE's by finding a global quantity
obeying a deterministic equation of the same type as the noiseless
ODE's (ordinary differential equations) of the model. For stochastic
systems that are either outright Gaussians (as for instance an
Ornstein-Uhlenbeck process), or that can be traced back to some
other Gaussian process (as a geometric Wiener process), this is
simple enough to accomplish because of both the linearity of the
involved SDE's and the symmetry of the distributions.

Take for instance the Gompertz stochastic model (for details see in
particular~\cite{ReportNoi}) satisfying the non-linear SDE
\begin{equation}\label{gompSDE}
    dX(t)=\big[X(t)-\alpha X(t)\,\ln{X(t)}\big]\,dt+X(t)\,dW(t)
\end{equation}
It is easy to see then that the transformed process $Y(t)=\ln{X(t)}$
satisfies the new, linear SDE
\begin{equation}\label{gomplinSDE}
    dY(t)=(1-D-\alpha Y(t))+dW(t)
\end{equation}
namely a modified Ornstein-Uhlenbeck equation with Gaussian
solutions: therefore the original process $X(t)$ has a log-normal
distribution. By taking the expectation of the linear
SDE\refeq{gomplinSDE} it is easy to see moreover that the averaged
quantity $\EXP{Y(t)}$ satisfies the ODE
\begin{equation}\label{gomplinODE}
    \frac{d\,\EXP{Y(t)}}{dt}=\big(1-D-\alpha\,\EXP{Y(t)}\big)
\end{equation}
Remark that it would not be expedient to directly take the
expectation of the SDE\refeq{gompSDE} because of its non linearity.
If instead we now consider the \emph{median} $\bm M[X(t)]$ of our
process it is possible to show that, because of the symmetry of the
Gaussian distribution of $Y(t)$, from the properties of the medians
we have
\begin{equation*}
    \bm M\left[X(t)\right]=\bm M\left[e^{Y(t)}\right]=e^{\bm M[Y(t)]}=e^{\EXP{Y(t)}}
\end{equation*}
and hence from\refeq{gomplinODE} it is easy to check that the median
satisfies the ODE
\begin{equation}\label{gompODE}
    \frac{d\,\bm M[X(t)]}{dt}=\bm M[X(t)]\big(1-D-\alpha\,\ln\bm M[X(t)]\big)
\end{equation}
that plays here the role of a coarse-grained ODE coinciding with a
slightly generalized Gomperts ODE
\begin{equation*}
    \dot{x}(t)=x(t)\big[\beta-\alpha\ln
    x(t)\big]\qquad\qquad\beta=1-D
\end{equation*}
and going back to its standard form\refeq{Geq2} for $D\to0$. This of
course also explains why the Gomperts process $X(t)$ (its
trajectories, distributions and moments) tends to its deterministic
behavior $x(t)$ when the noise is switched off.

Not so, instead, for the stochastic logistic instance because -- as
we have shown in the previous sections -- the distributions of the
solutions are much more tangled. We know indeed that its
trajectories, distributions and moments rightly show the bent to
converge toward their deterministic behavior for vanishing noise,
but in this case we are unable to recover a coarse-grained form of
the SDE by proceeding along the same way trod in the case the
Gompertz process. As a matter of fact the $\theta$-logistic
SDE\refeq{StochImplDef} can be reduced to linear coefficients
(see~\cite{ReportNoi}): with the transformation
$Y(t)=X^{-\theta}(t)$ we would in fact find
\begin{equation}\label{loglinSDE}
    dY(t)=\theta\big[1+((1+\theta)D-1)Y(t)\big]\,dt-\theta Y(t)\,dW(t)
\end{equation}
but, albeit possible, it would be useless to take its expectation
$\EXP{Y(t)}$. We know indeed that the path-wise solution of the
SDE\refeq{loglinSDE} is
\begin{equation*}
    Y(t)=e^{-\theta Z(t)}\left[Y_0+\theta\int_0^te^{\theta Z(u)}du\right]
\end{equation*}
where $Z(t)$ is defined in\refeq{Z}, and that its distributions
discussed in the Section \ref{pdm} are especially intricate,
confined on the positive half-axis and far from symmetric. As a
consequence, even if we can easily find an equation for
$\EXP{Y(t)}$, it would not be easy to manage a way to find a coarse
grained quantity of the process $X(t)$ obeying some form of its
noiseless equation as we did with the median in the Gompertz case,
and we plan to tackle this problem in our future inquiries.


\begin{thebibliography}{99}

\bibitem{muller} J. M\"uller and C. Kuttler, \emph{Methods and Models in Mathematical Biology},
Lecture Notes on Mathematical Modelling in the Life Sciences
(Springer-Verlag, Berlin Heidelberg, 2015); J. M\"uller,
\emph{Mathematical Models in Biology}, Lecture held in the
Winter-Semester 2003/2004 at the Centre for Mathematical Sciences,
Technical University Munich,

www.bionica.info/Biblioteca/Muller2004MathematicalModelsInBiology.pdf.

\bibitem{Ovaskainen} O. Ovaskainen and B. Meerson, Trends in ecology\& evolution 25, 643 (2010)

\bibitem{Salisbury} A. Salisbury, \emph{Mathematical Models in Population Dynamics},
Phd Thesis,

https://core.ac.uk/download/pdf/141995076.pdf

\bibitem{Murray1} J. D. Murray, \emph{Mathematical Biology I: An Introduction}
(Springer-Verlag, New York Berlin Heidelberg, 2002).

\bibitem{Murray2} J. D. Murray, \emph{Mathematical Biology II: Spatial Models and Biomedical Applications}
(Springer-Verlag, New York Berlin Heidelberg, 2003).

\bibitem{Verhulst} P.F. Verhulst, \emph{Notice sur la loi que la population suit
dans son accroissement}, Corr. Mat. et Phys. 10, 113–121 (1838);
P.F. Verhulst (1845) Nouveaux Memoires de l'Academie Royale des
Sciences et Belles-Lettres de Bruxelles 18, pp. 1-38.

\bibitem{Gompertz} B. Gompertz,
\emph{On the nature of the function expressive of the law of human
mortality, and on a new mode of determining the value of life
contingencies}, Phil. Trans. R. Soc. 115, 513 (1825).

\bibitem{Richards} F. J. Richards, "A Flexible Growth Function for Empirical Use",
Journal of Experimental Botany. 10 (2): 290–300 (1959).

\bibitem{GilpinAyala} M. E. Gilpin and F. J. Ayala,
\emph{Global Models of Growth and Competition}, PNAS, 70,
3590–3593 (1973).

\bibitem{Bellomo1} N. Bellomo, E. De Angelis and L. Preziosi,
\emph{Multiscale Modeling and Mathematical Problems Related to Tumor
Evolution and Medical Therapy}, Journal of Theoretical Medicine,
5(2), 111–136 (2003).

\bibitem{Bellomo2} N. Bellomo, N. K. Li and P. K. Maini, \emph{On the foundations of cancer modelling:
selected topics, speculations, and perspectives}, Mathematical
Models and Methods in Applied Sciences 18, 593–646 (2008).

\bibitem{Lowengrub} J. S. Lowengrub, H. B. Frieboes, F. Jin, Y-L. Chuang, X. Li, P. Macklin, S. M. Wise, and
V. Cristini, \emph{Nonlinear modelling of cancer: bridging the gap
between cells and tumours}, Nonlinearity 23(1), R1–R9 (2010).

\bibitem{Drasdo} D. Drasdo, S. Hoehme, and M. Block,
\emph{On the Role of Physics in the Growth and Pattern Formation of
Multi-Cellular Systems: What can we Learn from Individual-Cell Based
Models?}, J Stat Phys (2007) 128: 287.
https://doi.org/10.1007/s10955-007-9289-x.

\bibitem{Alekseev} O. Alekseev and M. Mineev-Weinstein,
\emph{Statistical mechanics of stochastic growth phenomena}, Phys.
Rev. E 96, 010103(R) (2017).

\bibitem{West} J. West and P. K. Newton, \emph{Cellular cooperation shapes tumor growth:
a statistical mechanics mathematical model}, bioRxiv preprint first
posted online Mar. 8, 2018, http://dx.doi.org/10.1101/278614.

\bibitem{Riffi} M. I. Riffi,
\emph{A Generalized Transmuted Gompertz-Makeham Distribution},
Journal of Scientific and Engineering Research, 5(8), 252-266
(2018).

\bibitem{Yamano} T. Yamano,
\emph{Statistical Ensemble Theory of Gompertz Growth Model}, Entropy
11, 807-819 (2009).

\bibitem{Wrycza} T. F. Wrycza,
\emph{Entropy of the Gompertz-Makeham mortality model}, DEMOGRAPHIC
RESEARCH, 30, 1397–1404 (2014).

\bibitem{Lande} R. Lande, S. Engen, and B.-E. Saether,
\emph{Stochastic population dynamics in ecology and conservation}
(Oxford University Press, 2003)

\bibitem{Gutierrez} R. Gutierrez-Jaimez, P. Roman, D. Romero, J.J. Serrano, F. Torres,
\emph{A new Gompertz-type diffusion process with application to
random growth}, Math. Biosci. 208, 147 (2007).

\bibitem{Schurz}
H. Schurz, \emph{Modeling, analysis and discretization of stochastic
logistic equations}, International journal of numerical analysis and
modeling, 4, 178-197 (2007).

\bibitem{Skiadas} C. H. Skiadas, \emph{Exact Solutions of Stochastic Differential Equations:
Gompertz, Generalized Logistic and Revised Exponential}, Methodol
Comput Appl Probab 12, 261–270 (2010).

\bibitem{Khodabin} M. Khodabin and N. Kiaee,
\emph{Stochastic Dynamical Theta-Logistic Population Growth Model},
SOP TRANSACTIONS ON STATISTICS AND ANALYSIS, 1, 1 (2014).

\bibitem{noi} S. De Martino and S. De Siena, \emph{Stochastic roots of growth phenomena},
Physica A 401, 207–213 (2014).

\bibitem{Yor1} M. Yor, \emph{Exponential Functionals of Brownian Motion and Related
Processes} (Springer, Berlin 2001).

\bibitem{Yor2} H. Matsumoto and M. Yor,
\emph{Exponential functionals of Brownian motion, I: Probability
laws at Fixed time}, Probability Surveys Vol. 2 (2005) 312-347.

\bibitem{Yor3} H. Matsumoto and M. Yor,
\emph{Exponential functionals of Brownian motion, II: Some related
diffusion processes}, Probability Surveys Vol. 2 (2005) 348–384.

\bibitem{Ashcroft} N. W. Ashcroft and N. D. Mermin, \emph{Solid State Physics}
(Arcourt College Publishers, 1976).

\bibitem{Bartlett} Bartlett, M.S., Gower, J.S., Leslie, P.H.,
\emph{A comparison of theoretical andempirical results for some
stochastic population models}, Biometrika 47, 1–11 (1960).

\bibitem{Ovaskainen2} O. Ovaskainen, \emph{The quasistationary distribution of the stochastic logistic model},
J. Appl. Prob. 38, 898-907 (2001).

\bibitem{Pasquali} S. Pasquali,
\emph{The stochastic logistic equation : stationary solutions and
their stability},Rendiconti del Seminario Matematico della
Universit\'a di Padova, tome 106, p. 165-183 (2001).

\bibitem{Nasell1} I. Nasell,
\emph{Extinction and quasi-stationarity in the Verhulst logistic
model}, J. Theor. Biol. 211, 11–27 (2001).

\bibitem{Nasell2}  I. Nasell, \emph{Extinction and quasi-stationarity in the Verhulst logistic model II},
www.math.kth.se/~ingemar/forsk/verhulst/verhulst.html.

\bibitem{Nasell3} I. Nasell,
\emph{Moment closure and the stochastic logistic model}, Theoretical
Population Biology 63, 159–168 (2003).

\bibitem{Ramasubramanian} B. Ramasubramanian,
\emph{Stochastic Differential Equations in Population Dynamics:
Numerical Analysis, Stability and Theoretical Perspectives},

https://pdfs.semanticscholar.org/cfe9/be4bf6e638b29b8c723cd3ba6d06225e1f48.pdf.

\bibitem{Tenkes} L.-M. Tenk\'es, R. Hollerbach, and E. Kim,
\emph{Time-dependent probability density functions and information
geometry in stochastic logistic and Gompertz models}, Journal of
Statistical Mechanics: Theory and Experiment 17, 123201 (2017).

\bibitem{ReportNoi} N. Cufaro Petroni, S. De Martino, and S. De Siena,
\emph{Gompertz and logistic stochastic dynamics: Advances in an
ngoing quest}, arXiv:2002.06409 [math.PR].

\bibitem{Gardiner} C. W. Gardiner, \emph{Handbook of Stochastic Methods}
(Springer-Verlag Berlin Heidelberg New York, 1994).

\bibitem{Gihman} I. I. Gihman and A.V. Skorohod,
\emph{Stochastic Differential Equations} (Springer, Berlin 1972).

\end{thebibliography}
\end{document}